\documentclass[aps,prb,twocolumn,showpacs,showkeys,footinbib,superscriptaddress]{revtex4-1}

\usepackage{amsmath}
\usepackage{amssymb}
\usepackage{graphicx}
\usepackage{bm}
\usepackage{color}
\usepackage{relsize}
\usepackage{braket}
\usepackage[caption=false]{subfig}

\usepackage{hyperref}
\usepackage[all]{hypcap}

\begin{document}
\title{Probing Many-Body Interactions in Monolayer Transition-Metal Dichalcogenides}
\date{\today}

\author{Dinh Van Tuan}
\affiliation{Department of Electrical and Computer Engineering, University of Rochester, Rochester, New York 14627, USA}
\author{Benedikt Scharf}
\affiliation{Department of Physics, University at Buffalo, State University of New York, Buffalo, NY 14260, USA}
\affiliation{Institute for Theoretical Physics, University of Regensburg, 93040 Regensburg, Germany}
\affiliation{Institute for Theoretical Physics and Astrophysics, University of W\"{u}rzburg, Am Hubland, 97074 W\"{u}rzburg, Germany}
\author{Zefang Wang}
\affiliation{Department of Physics and Center for Two-Dimensional and Layered Materials, The Pennsylvania State University, University Park, Pennsylvania 16802-6300, USA}
\author{Jie Shan}
\affiliation{Department of Physics and Center for Two-Dimensional and Layered Materials, The Pennsylvania State University, University Park, Pennsylvania 16802-6300, USA}
\author{Kin Fai Mak}
\affiliation{Department of Physics and Center for Two-Dimensional and Layered Materials, The Pennsylvania State University, University Park, Pennsylvania 16802-6300, USA}
\author{Igor \v{Z}uti\'c}
\affiliation{Department of Physics, University at Buffalo, State University of New York, Buffalo, NY 14260, USA}
\author{Hanan~Dery}
\email[]{hanan.dery@rochester.edu}
\affiliation{Department of Electrical and Computer Engineering, University of Rochester, Rochester, New York 14627, USA}
\affiliation{Department of Physics and Astronomy, University of Rochester, Rochester, New York 14627, USA}

\begin{abstract} 
Many-body interactions in monolayer transition-metal dichalcogenides are strongly affected by their unique band structure. We study these interactions by measuring the energy shift of neutral excitons (bound electron-hole pairs) in gated WSe$_2$ and MoSe$_2$. Surprisingly, while the blueshift of the neutral exciton, $X^0$, in electron-doped samples can be more than 10~meV, the blueshift in hole-doped samples is nearly absent. Taking into account dynamical screening and local-field effects, we present a transparent and analytical model that elucidates the crucial role played by intervalley plasmons in electron-doped conditions. The energy shift of $X^0$ as a function of charge density is computed showing agreement with experiment, where the renormalization of $X^0$ by intervalley plasmons yields a stronger blueshift in MoSe$_2$ than in WSe$_2$ due to differences in their band ordering.
\end{abstract}

\pacs{}
\keywords{}

\maketitle

\section{Introduction}

Monolayer transition-metal dichalcogenides (ML-TMDs) offer unique opportunities to test many-body interactions through changes in the charge density.\cite{Ye_Science12,Mak_NatMater13,Jones_NatNano13,Ross_NatComm13,Newaz_SolidStateComm13,Ugeda_NatMater14,Ganchev_PRL15,Chernikov_PRL15,Plechinger_PSS15,Shang_ACSNano15,Jones_NatPhys16,Hanbicki_AIP16} Their two-dimensional (2D) character and reduced screening enable the formation of tightly-bound excitons,\cite{Mak_PRL10,Splendiani_NanoLett10,Korn_APL11,Wang_NatNano12,Cheiwchanchamnangij_PRB12,Ramasubramaniam_PRB12,Komsa_PRB12,Xiao_PRL12,Berkelbach_PRB13a15,qiu_PRL13,Britnell_Science13,Song_PRL13,He_PRL14,Xu_NatPhys14,Chernikov_PRL14,Zhang_PRB14,Wu_PRB15,Li_PRB15,Zhang_NanoLett15,Robert_PRB16,Stier_JVSTB16,Mak_NatPhot16,Gunlycke_PCCP16} whose response to electrostatic doping provides valuable information on the Coulomb interactions of few-particle complexes,\cite{Chen_NatComm18,Ye_NatComm18,Li_NatComm18,Stevens_NatComm18,Barbone_NatComm18} or many-body effects when excitons interact with the background charge.\cite{Scharf_arXiv18,VanTuan_PRX17,Sidler_NatPhys17,Efimkin_PRB17,Schmidt_NanoLett16,Steinhoff_NanoLett14,Steinhoff_NatCom17,Steinhoff_PRB18} The dependence of the spectral position of the neutral exciton, $X^0$, on the gate-induced charge density is usually governed by two competing effects: Screening and band-gap renormalization (BGR).\cite{HaugKoch_Book,Haug_SchmittRink_PqE84,SchmittRink_PRB86,Rohlfing_PRB00} The background charge screens the electron-hole interaction of photoexcited bound pairs, thereby reducing the binding energy and causing  $X^0$ to blueshift towards the continuum of free electron-hole pairs. On the other hand, Coulomb exchange and correlation interactions between gate-induced charges shrink the band-gap energy and redshift the overall optical spectrum. Because long-wavelength charge excitations (intravalley plasmons) dominate both screening and BGR, the two effects almost completely compensate each other and the overall outcome is a nearly fixed spectral position of $X^0$. 

The above description is common in conventional semiconductors and can be modeled by a quasistatic Bethe-Salpeter Equation (BSE).\cite{HaugKoch_Book,Haug_SchmittRink_PqE84} However, it cannot explain why the blueshift of $X^0$ is much stronger for electron-doped ML-TMDs compared with hole-doped ones. In fact, there are two compelling reasons that long-wavelength charge excitations should yield similar rather than different energy shifts in the two doping cases. The first reason is that the electron and hole effective masses are similar and the second one is that neither the conduction nor valence band is degenerate.  Accordingly, all that the long-wavelength charge excitations can explain in ML-TMDs is the BGR and the eventual merging of the exciton into the continuum at elevated charge densities.\cite{Scharf_arXiv18}

\begin{figure}
\centering
\includegraphics*[width=9cm]{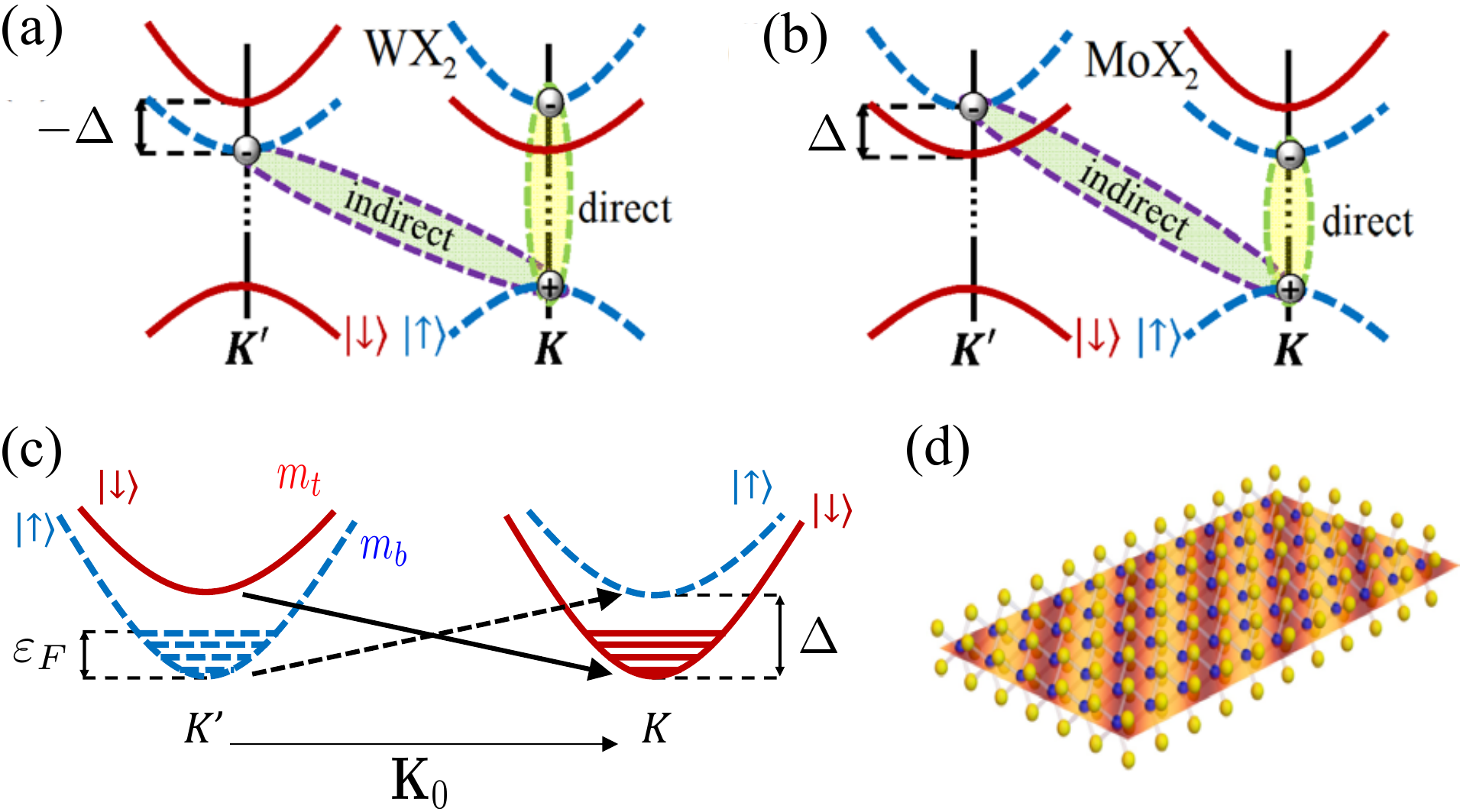}
\caption{(a,b) Low-energy band structure around the $\bm{K}/\bm{K}'$ points for WX$_2$ and MoX$_2$ MLs, respectively, where X denotes S or Se. Direct and indirect excitons are shown, where the spin of the bands is color coded. $|\Delta|$ is the conduction-band spin-splitting energy. (c) The intervalley Coulomb interaction in ML-TMDs. Spin-conserving charge excitations from the $K'$ to the $K$ valleys. $\varepsilon_F$ is the Fermi energy. (d) The resulting shortwave charge fluctuations in the monolayer.}\label{fig:cartoon}
\end{figure}

In this work, we first experimentally quantify the blueshift of excitons in ML-MoSe$_2$ and ML-WSe$_2$ by optical reflectance spectroscopy.  We then present an analytical model that quantifies the coupling between low-energy exciton states and intervalley plasmons in ML-TMDs (Fig.~\ref{fig:cartoon}).\cite{Dery_PRB16,Groenewald_PRB16} One advantage of the theory is that it readily calculates many-body interactions in the exciton spectrum without the need to invoke a computationally intensive dynamical BSE model.\cite{VanTuan_PRX17} Most importantly, the interaction between the exciton and the shortwave plasmons is shown to explain on equal footing both the blueshift of $X^0$ in electron-doped MLs and the emergence of the optical sideband in electron-doped tungsten-based  MLs. The theory captures the observations that the blueshift of $X^0$ is stronger in ML-MoSe$_2$, that it is absent in hole-doped ML-TMDs, and that the optical sideband neither emerges in hole-doped MLs nor in electron-doped molybdenum-based MLs.  

This paper is organized as follows. We first present the experimental results (Sec.~\ref{sec:exp}), followed by a detailed analysis of the theoretical model in Sec.~\ref{sec:theory}. We present results that can be directly compared to our experimental results in  Sec.~\ref{sec:res_conc}, which also concludes this work. Appendix~\ref{app:bgr} includes technical details of the calculation of the BGR.

\section{Experiment}\label{sec:exp}

\begin{figure}[hbtp]
\centering
\includegraphics[width=6cm]{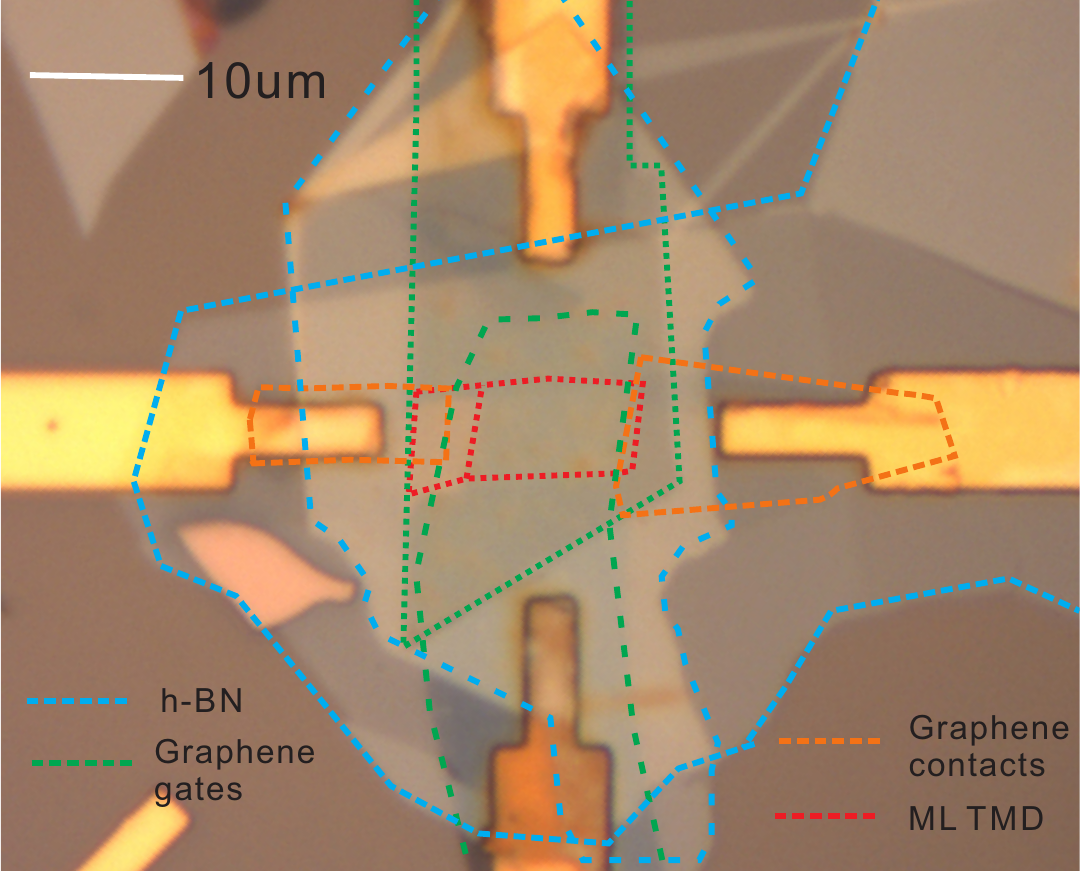}
\caption{Optical microscope image of a dual-gate device of WSe$_2$. The boundary of each component is shown in dashed lines. The scale bar is 10 $\mu$m.}\label{fig:Device}
\end{figure}

We measure the evolution of the exciton spectra in ML-MoSe$_2$ and ML-WSe$_2$ as a function of the gate-induced electrostatic doping through reflection contrast measurements performed on dual-gate  field-effect transistors. The devices were fabricated  by the dry transfer technique, making use of $\sim$20-nm-thick hexagonal boron nitride (h-BN) as top and back gate dielectric. 
\cite{Wang_NanoLett17,Wang_NatNano17} Few-layer graphene is used for both top and back gate electrodes. Few-layer graphene is also used for source and drain contacts to monolayer WSe$_2$ (MoSe$_2$). Atomically thin flakes of h-BN, graphene, and WSe$_2$ (MoSe$_2$) were first mechanically exfoliated from bulk crystals onto silicon substrates covered with a 280-nm layer of thermal oxide. Their thickness was first estimated from the optical contrast and then confirmed by the atomic force microscopy or photoluminescence spectroscopy. The chosen flakes were then picked up layer by layer with a stamp made of a thin layer of polypropylene carbonate (PPC) on polydimethylsiloxane. Using a micromanipulator under a microscope, we were able to align the flakes with the accuracy of $\sim$1 $\mu$m. The stack was then released onto a silicon substrate with pre-patterned gold electrodes to form the dual gate field-effect transistors. The PPC residue on the device was removed before the optical measurements by dissolving it in anisole. Figure~\ref{fig:Device} shows an optical microscope image of a device in which ML-WSe$_2$ serves as the active layer. 

The reflection contrast measurement was performed in a close-cycle cryostat from 4 to 300~K. Broadband radiation from a supercontinuum light source was focused by a 40x objective onto the sample to a spot diameter of $\sim$1~$\mu$m. The reflected light was collected by the same objective and detected by a spectrometer equipped with a charge-coupled-device (CCD). The excitation power on the device was kept below 10 $\mu$W. The reflection contrast spectrum $\delta R/R$ was obtained by measuring the reflectance from the part of the device with and without ML-WSe$_2$ (ML-MoSe$_2$).

With the combination of the top and back gates, the doping density and the vertical electric field in monolayer WSe$_2$ (MoSe$_2$) can be tuned independently. We focus on the doping density effects in this study. The vertical electric field was kept at zero by applying the same voltage on both the top and back gate since the top and back h-BN dielectric layer have the same thickness. The doping density (including both the free and localized charge carriers) can be evaluated by $n=\epsilon\epsilon_0 V / e t$, where $e=1.6\times10^{-19}$~C is the elementary charge, $\epsilon_0=8.85\times10^{-14}$~F/cm is the vacuum permittivity, and $\varepsilon$ is the relative dielectric constant of h-BN. The latter is found from the in-plane and out-of-plane components according to, $\epsilon = \sqrt{\epsilon_{\parallel}\epsilon_{\infty}}$, and it becomes $\sim$3.8 in the high-frequency regime and $\sim$4.9 in the static limit.\cite{VanTuan_PRB18} The thickness of the h-BN layer is $t \sim 20$~nm, and $V$ is the combined top and back gate voltage. We then get that 1~V is equivalent to a doping density of $\sim10^{12}$~cm$^{-2}$.
 
\begin{figure*}
\centering
\includegraphics*[width=17cm]{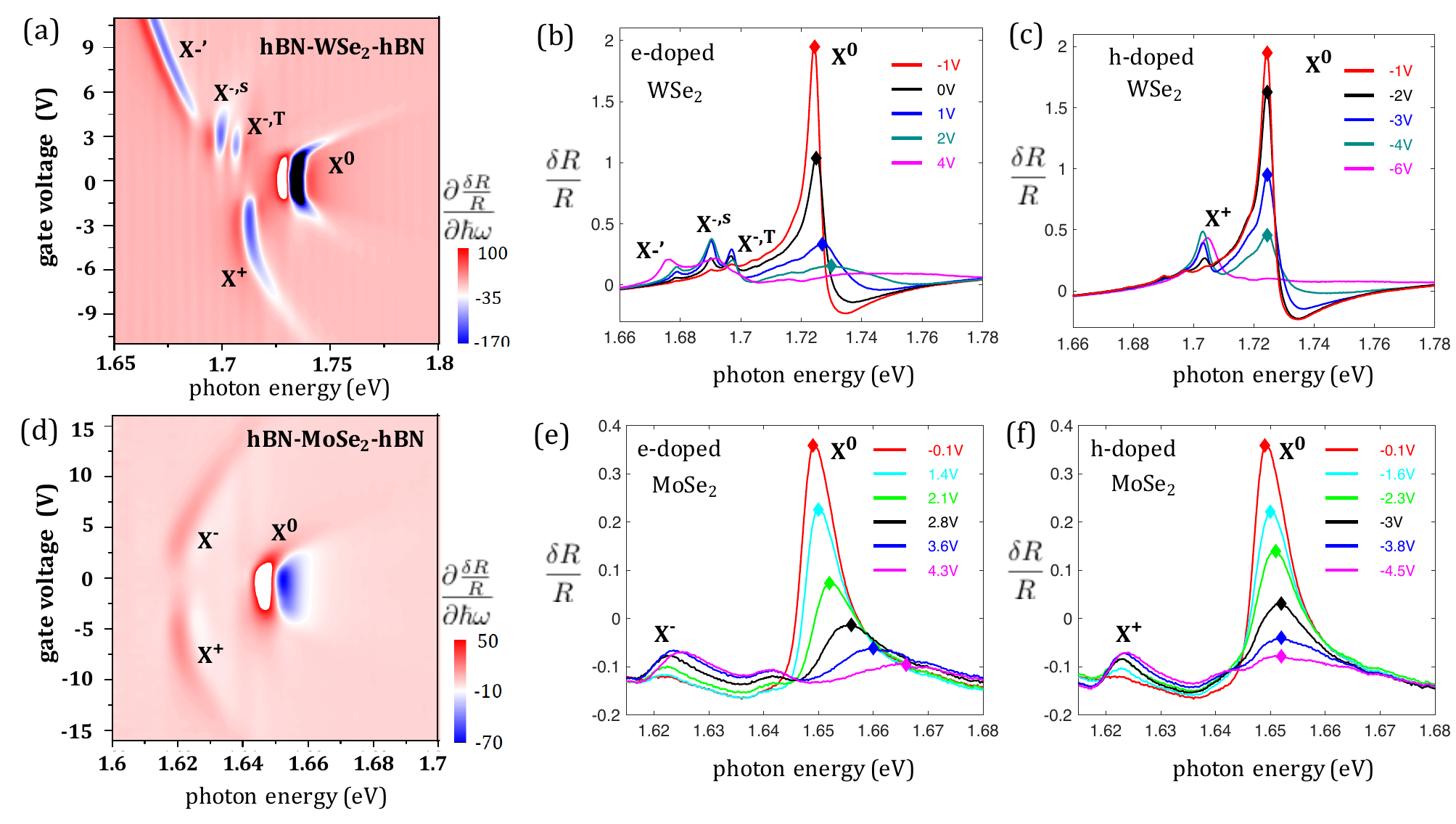}
\caption{Measured blueshift of $X^0$ in gated ML-TMDs: (a) A colormap of the energy derivative of the reflectance contrast spectra ($\partial (\delta R/R)/ \partial \hbar \omega$) at 4K in WSe$_2$. (b,c) $\delta R/R$ of electron-doped and hole-doped cases for different gate voltages. The diamond symbols show the peak position of $X^0$. (d,e,f) show the respective measured results but for MoSe$_2$.}\label{fig:Exp}
\end{figure*}

\subsection{Results}

Figure~\ref{fig:Exp} shows the measured reflectance contrast spectra in the energy range of $X^0$ in gated ML-WSe$_2$ and ML-MoSe$_2$. To increase the contrast for the resonance features, the derivative of the reflectance contrast is shown in Figs.~\ref{fig:Exp}(a) and (d). The measurements clearly show that $X^0$ exhibits a blueshift with electron doping [positive gate voltage; Figs.~\ref{fig:Exp}(b,e)], where the shift is $\sim$20~meV for the shown range of gate voltages in electron-doped ML-MoSe$_2$ and  $\sim$5~meV in electron-doped ML-WSe$_2$. Hole-doped samples, on the other hand, exhibit no or at most a tiny blueshift, while $X^0$ decays with increasing hole doping [Figs.~\ref{fig:Exp}(c,f)]. In addition to $X^0$, Fig.~\ref{fig:Exp} also shows positively and negatively charged excitons, $X^{\pm}$ or their singlet and triplet spin configurations, $X^{-,S}$ and $X^{-,T}$, in electron-doped ML-WSe$_2$.\cite{Jones_NatPhys16,Courtade_PRB16,Plechinger_NatCom16} Also shown is the optical sideband, $X-$', that we have recently associated to the unique coupling of neutral excitons and intervalley plasmons in W-based compounds.\cite{VanTuan_PRX17} 

\section{Theory}\label{sec:theory}

We focus on the behavior of neutral excitons as a function of the background charge density in the ML. The $X^0$ peak in ML-TMDs originates from bright direct $1s$ excitons, which mainly arise from the optical transition between the topmost spin-split valence band and the conduction band with the same spin and valley quantum numbers.\cite{Xu_NatPhys14} As shown in Figs.~\ref{fig:cartoon}(a,b), Mo- and W-based compounds are different in that direct optical transitions in the former (latter) involve the bottom (top) spin-split valleys of the conduction band.\cite{Song_PRL13,Kormanyos_PRX14,Dery_PRB15,Zhang_PRL15,Withers_NanoLett15,Wang_NatComm15,Arora_Nanoscale15} Therefore, the direct-exciton mass is $M_d = m_{ct} +m_{vt}$ for WSe$_2$ or $M_d = m_{cb} +m_{vt}$ for MoSe$_2$, where $m_{ct}$($m_{cb}$) denotes the electron effective mass in the top (bottom) valley of the conduction band, and $m_{vt}$ is the hole effective mass in the top spin-split valence band. Conversely, the mass of the indirect exciton is $M_i = m_{cb} +m_{vt}$ for WSe$_2$ or $M_i = m_{ct} +m_{vt}$ for MoSe$_2$, as shown in Figs.~\ref{fig:cartoon}(a,b).

The behavior of $X^0$ is studied from the relation between absorption of a photon with energy $\hbar \omega$ and the direct-exciton Green's function,\cite{Haug_SchmittRink_PqE84}
\begin{equation}\label{Eq:alpha}
\alpha(\hbar \omega) \propto -\mathrm{Im}\left[{G}_d(\bm{q}=\bm{0}, \hbar \omega - E_\mathrm{g,d})\right].
\end{equation}
 $\bm{q}$ is the exciton's center-of-mass wavevector, where the limit $q \rightarrow 0$ applies for excitons in the light cone. $E_\mathrm{g,d}$ is the band-gap energy between the valence- and conduction-band valleys from which the direct exciton arises. The Green's function reads 
\begin{equation}\label{Eq:G}
G_d(\bm{q},E)= \left[ E -E_{d,\bm{q}}-\Sigma_\mathrm{s}(\bm{q},E) + i \Gamma(E) \right]^{-1},
\end{equation}
where $\Gamma$ denotes broadening, $\Sigma_\mathrm{s}$ is a self-energy correction to be discussed later, and
\begin{equation}\label{Eq:D_ENG}
E_{d,\bm{q}}= E_{d} + \hbar^2q^2/2M_d.
\end{equation}
$E_{d}$ is the direct-exciton energy level below the continuum (i.e., $|E_{d}|$ is its binding energy). The pole of $G_d(\mathbf{q}=0,E)$ is at $\hbar \omega = E_\mathrm{g,d} + E_{d} + \Sigma_\mathrm{s}$. The sum $E_\mathrm{g,d} + E_{d}$ is largely unaffected when the charge density in the ML increases because of the offset between shrinkage of the band gap and smaller binding energy due to screening.\cite{Haug_SchmittRink_PqE84} 

The blueshift in the absorption spectrum mostly arises from the exciton's self-energy, $\Sigma_\mathrm{s}$. We consider the self-energy correction from virtual transitions between direct and indirect excitons mediated by shortwave (intervalley) plasmons.\cite{VanTuan_PRX17} The plasmon wavevector, $\bm{K}_0 + \bar{\bm{q}}$, is the sum of a small component $\bar{\bm{q}}$, and the large central wavevector $\bm{K}_0$ that connects the centers of the time-reversed valleys ($K_0 = 4\pi/3a$ where $a\simeq 3.2$~\AA~ is the triangular lattice constant). Using the finite-temperature Green's function formalism,\cite{Mahan_BookChapter} the self-energy of direct excitons due to shortwave plasmons follows from
\begin{eqnarray}\label{Eq:SE_Omega}
\Sigma_{s}(\bm{q},\Omega)=-k_\mathrm{B}T \! \! \sum_{\bar{\bm{q}},\Omega'}\! \left| \mathcal{M}_{\bar{\bm{q}}}\right|^2 \! D(\Omega \!-\! \Omega',\bar{\bm{q}}) G_{i}(\bar{\bm{q}}\!+\! \bm{q},\Omega'). \,\,\,\,\,\,\,
\end{eqnarray}
$k_BT$ is the thermal energy, and $\Omega, \Omega'$ denote even (boson) imaginary Matsubara energies that will eventually be analytically continued into the real-energy axis ($\Omega \rightarrow E + i\Gamma_{\Sigma}$).  The sum over $\bar{\bm{q}}$ is restricted to the range of damping-free plasmon propagation range.\cite{VanTuan_arXiv19} $D(\Omega,\bar{\bm{q}})$ is the  intervalley-plasmon propagator
\begin{eqnarray}\label{eq:D}
D(\Omega,\bar{\bm{q}})= \frac{2\hbar \omega_{\bar{\bm{q}}}}{\Omega^2- \hbar^2 \omega^2_{\bar{\bm{q}}}} \,,\,\,\,\,\,\,
\end{eqnarray}
where $\omega_{\bar{\bm{q}}}$ is the collective excitation frequency to be defined in Sec.~\ref{sec:wM} along with the  exciton-plasmon interaction matrix element, $\mathcal{M}_{\bar{\bm{q}}}$.  Finally, $G_{i}(\bar{\bm{q}},\Omega)$ is the unperturbed indirect-exciton Green's function (prior to renormalization by intervalley plasmons), 
\begin{eqnarray}\label{eq:Gi}
G_{i}(\bar{\bm{q}},\Omega) = \frac{1}{\Omega-E_{i,\bar{\bm{q}}}} \,.\,\,\,\,\,\,
\end{eqnarray}
The energy $E_{i,\bar{\bm{q}}}= E_{i} + \hbar^2\bar{q}^2/2M_i$ is defined similarly to $E_{d,\bm{q}}$ in Eq.~(\ref{Eq:D_ENG}), but with indirect exciton parameters. 

The self-energy computation is greatly simplified by using the approximated form of $G_{i}(\bar{\bm{q}},\Omega)$ in Eq.~(\ref{eq:Gi}) instead of calculating its values from an intensive dynamical BSE model.\cite{VanTuan_PRX17} The summation over $\Omega'$ in Eq.~(\ref{Eq:SE_Omega}) can be transformed into contour integration in the complex plane by using the identity
\begin{eqnarray}\label{Eq:idenitity}
ik_BT \sum_{\Omega'} F(\Omega') = \oint_C \frac{d\Omega'}{2\pi} \frac{F(\Omega')}{e^{\Omega'/k_BT} - 1 } \,,
\end{eqnarray}
where the contour encircles the poles of $F(z)$ in the positive sense. Considering direct excitons in the light cone [$\bm{q} \rightarrow 0$ in Eqs.~(\ref{Eq:alpha}) and (\ref{Eq:SE_Omega})], we then get that 
\begin{eqnarray}\label{Eq:SE_supp}
\Sigma_\mathrm{s}(\Omega) = \sum\limits_{\bar{\bm{q}}}\left|\mathcal{M}_{\bar{\bm{q}}}\right|^2 && \left[\frac{g(E_{i}(\bar{\bm{q}}))-g(\hbar \omega_{\bar{\bm{q}}})}{\Omega+ \hbar \omega_{\bar{\bm{q}}} - E_{i}(\bar{\bm{q}}) } \right. \nonumber \\ && -  \left. \frac{g(E_{i}(\bar{\bm{q}}))-g(-\hbar \omega_{\bar{\bm{q}}})}{\Omega-  \hbar \omega_{\bar{\bm{q}}}  -E_{i}(\bar{\bm{q}})}\right] . \,\,\,\,\,\,\,\,\,
\end{eqnarray} 
Noting that $E_{i}(\bar{\bm{q}})$ is negative, the low-temperature Bose-Einstein distributions follow $g(E_{i}(\bar{\bm{q}})) \rightarrow -1 $, $g(\hbar \omega_{\bar{\bm{q}}}) \rightarrow 0$, and $g(-\hbar \omega_{\bar{\bm{q}}}) \rightarrow -1$. Using these limits, we finally arrive at
\begin{eqnarray}\label{Eq:SE}
\Sigma_\mathrm{s}(\Omega)= - \sum\limits_{\bar{\bm{q}}} \frac{\left|\mathcal{M}_{\bar{\bm{q}}}\right|^2}{\Omega+ \hbar \omega_{\bar{\bm{q}}}  - E_{i,\bar{\bm{q}}}}.
\end{eqnarray}

\subsection{Intervalley plasmons in ML-TMDs} \label{sec:wM}

Both the exciton-plasmon matrix element, $\mathcal{M}_{\bar{\bm{q}}}$, and the collective excitation frequency, $\omega_{\bar{\bm{q}}}$, are related to the dynamical Coulomb potential in its shortwave limit. We provide a brief summary below and refer interested readers to Ref.~[\onlinecite{VanTuan_arXiv19}]  wherein a comprehensive analysis of intervalley plasmons can be found. The dynamical Coulomb potential,
\begin{equation}
W(\bm{q},\omega) =  \frac{V_{\bm{q}}}{\epsilon(\bm{q},\omega)}  \,, \label{eq:eps_dyn}
\end{equation}  
is expressed through the bare Coulomb potential, $V_{\bm{q}}$, and the dynamical dielectric function $\epsilon(\bm{q},\omega)$.  Plasmons are found from  the solution of $\epsilon(\bm{q},\omega)=0$. The damping-free propagation range,  $q< q_{\text{max}}$, is defined by solutions with real-value plasmon frequency for a given ${\bm{q}}$.  Focusing on the shortwave limit, $\bm{q}=\bm{K}_0+\bar{\bm{q}}$ where $K_0 \gg \bar{q}$, and making use of the single-plasmon pole (SPP) approximation, the excitation spectrum of the dynamical dielectric function is replaced by a single collective excitation frequency,\cite{VanTuan_arXiv19}
\begin{equation}
\frac{V_{\bm{q}}}{{\epsilon}({\bm{q}},\omega)} \simeq  V_{\bm{K}_0} \left( 1 +  \frac{r(\bar{\bm{q}})}{\omega^2-\omega^2_{\bar{\bm{q}}} }  \right).\,\, \label{eq:spp}
\end{equation}  
We have used the fact that $V_{\bm{q}} \simeq V_{\bm{K}_0}$ for the bare Coulomb potential ($K_0 \gg \bar{q}$). The residue, $r(\bar{\bm{q}})$, is found from the conductivity sum rule, or equivalently, from the asymptotic behavior of the dynamical dielectric function at high-frequencies, $\omega^2 \gg \omega^2_{\bar{\bm{q}}}$. When the zero-temperature dynamical dielectric function under the random-phase approximation (RPA) is replaced with the SPP form, we find that \cite{VanTuan_arXiv19}
 \begin{eqnarray}
r_{s}(\bar{\bm{q}})= \frac{2\alpha_0  \varepsilon_F}{\hbar^2} && \left[ (1-c_0)\Delta_c + \left( 1 + \frac{c_0}{1+\beta} \right) \varepsilon_{t,\bar{\bm{q}}} \right.  \nonumber \\ && \left. + \frac{\beta(1-c_0^2)}{2} \varepsilon_F \right] .\,
  \label{eq:rs}
  \end{eqnarray}
Using Fig.~\ref{fig:cartoon}(c) as a guide, we first explain the meaning of these parameters. $\varepsilon_F$ is  the Fermi energy measured from the edge of the bottom valley in the conduction band and $\Delta_c$ is the spin-splitting energy between the bottom and top valleys. $\varepsilon_{t,\bar{\bm{q}}} = \hbar^2 \bar{q}^2 / 2m_{ct}$ is the kinetic energy in the top valley, and $\beta=m_{cb}/m_{ct}-1$ is the valley mass asymmetry between the bottom and top valleys. $c_0 = 0$ when $\varepsilon_F < \Delta_c$ or $c_0= (\varepsilon_F - \Delta_c)/(\beta+1)\varepsilon_F$ when $\varepsilon_F > \Delta_c$. 

The parameter $\alpha_0$ in Eq.~(\ref{eq:rs}) deserves special attention.  Intervalley plasmons can propagate without damping in the range $\bar{q} \leq q_{\mathrm{max}}$, where $q_{\mathrm{max}}$ is commensurate with both $\alpha_0$ and the charge density. As such, $\alpha_0$ is a measure for the importance of intervalley plasmons in a multi-valley 2D crystal. Their effect is measurable when $\alpha_0$ is comparable or larger than unity. This parameter is defined by\cite{VanTuan_arXiv19}
\begin{eqnarray}
\alpha_0 = \frac{m_{cb}}{\hbar^2  } \cdot \frac{AV_{\bm{K}_0}}{2\pi \eta_c}  =  \frac{m_{cb}}{\hbar^2  } \cdot \frac{ e^2}{  \eta_c K_0 \epsilon_d(K_0)}\, ,\label{eq:alpha0}
\end{eqnarray}
where $A$ is the sample area and $\epsilon_d(K_0)$ is the non-local dielectric constant at $q=K_0$.  The non-local dielectric function is not related to the static limit $\omega \rightarrow 0$ of the dynamical dielectric function. The role of the former is to capture the $q$-dependence of the effective dielectric constant due to material parameters of the ML and its surrounding. The dynamical dielectric function, on the other hand, describes the response of the delocalized electrons (or holes) in the ML to a test charge, and in the limit of zero charge density we get $\epsilon(\bm{q},\omega) \rightarrow 1$. 

Next, we discuss the local-field-effect parameter $\eta_c$ in Eq.~(\ref{eq:alpha0}). Its general form follows\cite{VanTuan_arXiv19}
\begin{eqnarray}
\frac{1}{\eta_c} =   \sum_{\mathbf{G}}  \frac{ V_{\mathbf{K}_0+\mathbf{G}} }{  V_{\mathbf{K}_0}  }   |\mathcal{F}_c(\mathbf{K}_0+\bm{G})|^2 \,,\,\,\,\,\, \label{eq:eta}
\end{eqnarray}
where the sum runs over reciprocal lattice vectors ($\bm{G}$), and 
\begin{eqnarray}
\mathcal{F}_c(\bm{K}_0+\bm{G}) =  \langle \mathbf{K}_c' | e^{i(\mathbf{K}_0 + \bm{G})\mathbf{r}} | \mathbf{K}_c \rangle. \,\,\,\,\, \label{eq:Fq}
\end{eqnarray}
$| \mathbf{K}_c \rangle$ and $| \mathbf{K}_c' \rangle$ are the conduction-band states at the valley center, governed by the orbital $d_{z^2}$ of the transition-metal atom. Local-field effects play an important role because the ratio $V_{\mathbf{K}_0+\mathbf{G}} /  V_{\mathbf{K}_0}$ in Eq.~(\ref{eq:eta}) is not negligible: $K_0$ is comparable to $| \mathbf{K}_0+\mathbf{G} |$ for the first few umklapp processes (when the amplitude of $G$ is comparable to that of the reciprocal lattice basis vectors). 

Having found the residue and explained the physical meaning of all of its related parameters, we can find the collective excitation frequency from the asymptotic behavior of the dynamical dielectric function at the static limit, $\omega^2 \ll \omega^2_{\bar{\bm{q}}}$.  When the dynamical dielectric function under the random-phase approximation (RPA) is replaced with the SPP form, we find that\cite{VanTuan_arXiv19}
  \begin{eqnarray}
  \omega^2_{\bar{\bm{q}} }  =  r(\bar{\bm{q}}) \left[ 1 + \frac{|\beta|}{2\alpha_0 \mathcal{G}_{\bar{\bm{q}}}}\right]  ,\label{eq:wqs}
   \end{eqnarray} 
where
 \begin{eqnarray}
\!\!\! \mathcal{G}_{\bar{\bm{q}}} =  \ln \frac{1 + |\beta|\mathcal{R}(\Delta_c+\varepsilon_{t,\bar{\bm{q}}},(1+\beta)\varepsilon_{t,\bar{\bm{q}}},\varepsilon_F) }{1 + |\beta|\Theta(\varepsilon_F-\Delta_c)\mathcal{R}(\Delta_c-\varepsilon_{b,\bar{\bm{q}}},\varepsilon_{b,\bar{\bm{q}}},c_0\varepsilon_F) }\,.\,\,\,\,\,\,\,
\label{eq:R}
   \end{eqnarray} 
$\Theta(\varepsilon_F - \Delta)$ is the step function and $\varepsilon_{b,\bar{\bm{q}}} = \hbar^2 \bar{q}^2 / 2m_{cb}$ is the kinetic energy in the bottom valley. Finally, 
\begin{eqnarray}
\mathcal{R}(\varepsilon_1,\varepsilon_2,\varepsilon_3) \!=\! \frac{\sqrt{(\varepsilon_1+\beta\varepsilon_3)^2 \!-\! 4\varepsilon_2\varepsilon_3} \!-\! (\varepsilon_1\!-\!|\beta|\varepsilon_3)}{(|\beta|+\beta)\varepsilon_1 - 2\varepsilon_2}   \,\,.\,\,\,\,\,\label{eq:R}
   \end{eqnarray} 

\subsubsection{The interaction of intervalley plasmons with excitons}

Similar to the case of $X$-ray catastrophe in metals, the electrons Fermi sea from which the plasmons emerge is shaken up during photoexcitation. We first discuss the interaction of intervalley plasmons with a test charge such as a remote electron that impinges on the crystal, followed by the changes needed to evaluate the interaction with excitons. Using second quantization and defining the plasmon creation and annihilation operators by $b^{\dagger}$ and $b$, the interaction between a test charge and intervalley plasmons reads\cite{VanTuan_arXiv19}
\begin{eqnarray} \label{eq:Hp}
\!\!\!\! H_p(\bm{r}) = \sum_{\bm{q}_G} V_{\bm{q}} \lambda_{\bar{\bm{q}}} \mathcal{F}(\bm{q})   \left( b_{-\bar{\bm{q}}}+b^{\dagger}_{\bar{\bm{q}}} \right) e^{-i{\bm{q}}\bm{r}}.\,\,\,
\end{eqnarray} 
where 
\begin{eqnarray}
\lambda_{\bar{\bm{q}}} =  \sqrt{  \frac{ A m_{cb} r(\bar{\bm{q}})}{4\pi \alpha_0 \hbar \omega_{\bar{\bm{q}}} }  }  .\,\,\,
\end{eqnarray} 
The sum in Eq.~(\ref{eq:Hp}) runs over  $\bm{q}_G = \bm{G} + \bar{\bm{q}}$. Here, $\bm{q}= \bm{K}_0 + \bm{q}_G$ can take values outside the first Brillouin zone because the test charge is not part of the Fermi sea of electrons from which the plasmons emerge. Alternatively, we say that the test charge can be everywhere in the crystal and not only in lattice sites.

The plasmon-exciton interaction is different from the interaction between a plasmon and a test charge because the wavefunctions of the electron and hole in the exciton are not plane waves but rather Bloch waves.  To derive the exciton-plasmon interaction matrix element, we assume that the short-range Coulomb interaction associated with intervalley plasmons does not allow for one charge in the exciton to screen the interaction of the opposite charge with the plasmon: $K_0a_X \sim  7.8$ where $a_X \sim 1$~nm is the exciton Bohr radius.\cite{Stier_NanoLett16} When the coupling is between direct and indirect excitons (Fig.~\ref{fig:cartoon}), the electron component of the exciton goes through a spin-conserving intervalley transition while the hole is a spectator.  The matrix-element reads
 \begin{eqnarray} \label{eq:Mq_general}
\mathcal{M}_{c,\bar{\bm{q}}} &=& \langle X_d, n_{\bar{\bm{q}}} \pm 1 | H_p(\bm{r}) | X_i(\bar{\bm{q}}), n_{\bar{\bm{q}}} \rangle \,\,\,,
\end{eqnarray}
where $X_{d}$ denotes a direct exciton in the light cone and $X_i(\bar{\bm{q}})$ denotes an indirect exciton. $n_{\bar{\bm{q}}}=\langle b^{\dagger}_{\bar{\bm{q}}} b_{\bar{\bm{q}}} \rangle$ denotes the plasmon number where the $+$ ($-$) sign is for plasmon emission (absorption).  To evaluate this matrix element, we consider the Bloch wave of the electron component in the exciton, 
 \begin{eqnarray} \label{eq:bloch}
\psi_e(\bm{k}) = \sqrt{\frac{1}{N}} \sum_j \exp(i \bm{k}\bm{R}_j) \phi_{\bm{k}}(\bm{r}-\bm{R}_j) \,\,\,.
\end{eqnarray}
$N$ is the number of unit cells, $\bm{R}_j$ are the lattice points, and $\phi_{\bm{k}}$ is the orbital composition of the state (governed by the orbital $d_{z^2}$). Considering a simple tight-binding model where the overlap between atomic orbitals of different lattice sites is neglected, the matrix element in Eq.~(\ref{eq:Mq_general}) becomes
 \begin{eqnarray}
\mathcal{M}_{c,\bar{\bm{q}}} =  \lambda_{\bar{\bm{q}}}  \sum_{\bm{G}} V_{\mathbf{K}_G} |\mathcal{F}_c(\mathbf{K}_G)|^2 =  \sqrt{  \frac{ \pi \alpha_0 \hbar^3}{ A m_{cb}} \frac{ r(\bar{\bm{q}})}{\omega_{\bar{\bm{q}}} }  }  ,\,\,\,\,\,\,\,\,\,\,\,
\label{eq:M_ex_e}
\end{eqnarray} 
where $\mathbf{K}_G = \mathbf{K}_0+\bm{G}$. We have made use of the facts that $\bar{q} \ll K_0$, and thus, $\mathcal{F}_c(\mathbf{q}) \simeq \mathcal{F}_c(\mathbf{K}_G)$ and $V_{\mathbf{q}} \simeq V_{\mathbf{K}_G}$.

\subsubsection{Hole-doped ML-TMDs}

The formalism and parameters so far assume electron-doped ML-TMDs. The case in hole-doped conditions is similar but with the following changes. Conduction-band subscripts are replaced by valence-band ones ($c \rightarrow v$). The index of the top and bottom valleys is exchanged ($b \leftrightarrow t$) because electrons first populate the bottom valleys in the conduction band, while holes populate the top valleys in the valence band. Finally, we use the orbital $d_{(x \pm iy)^2}$ to evaluate local field effects of the valence-band states instead of $d_{z^2}$.\cite{Zhu_PRB11}  The matrix element for the exciton-plasmon interaction becomes
 \begin{eqnarray}
\mathcal{M}_{v,\bar{\bm{q}}} =  \sqrt{  \frac{ \pi \widetilde{\alpha}_0 \hbar^3}{ A m_{vt}} \frac{ r(\bar{\bm{q}})}{\omega_{\bar{\bm{q}}} }  }  .\,\,\,\,\,\,\,\,\,\,\,
\label{eq:M_ex_h}
\end{eqnarray} 
The residue and single collective frequency in Eqs.~(\ref{eq:rs}) and (\ref{eq:wqs}) are now evaluated with valence band parameters, and 
\begin{eqnarray}
\widetilde{\alpha}_0 =  \frac{m_{vt}}{\hbar^2  } \cdot \frac{ e^2}{  \widetilde{\eta} K_0 \epsilon_d(K_0)}\, ,\label{eq:alpha0_prime}
\end{eqnarray}
where 
\begin{eqnarray}
\frac{1}{\widetilde{\eta}} =   \sum_{\mathbf{G}}  \frac{ V_{\mathbf{K}_0+\mathbf{G}} }{  V_{\mathbf{K}_0}  }   \mathcal{F}_v(\mathbf{K}_0+\bm{G}) \mathcal{F}_c(\mathbf{K}_0+\bm{G}) \,.\,\,\,\,\, \label{eq:eta_prime}
\end{eqnarray}
Here, the component  $\mathcal{F}_c(\mathbf{K}_0+\bm{G})$ stems from the electron component of the exciton when it goes through a spin-conserving intervalley transition, while  the component  $\mathcal{F}_v(\mathbf{K}_0+\bm{G})$ stems from the intervalley plasmon part that is now governed by the Fermi sea of holes. We note that a term $|\mathcal{F}_v(\mathbf{K}_0+\bm{G})|^2$ is expected in a Fermi sea of holes when the scattering is between type-A and type-B excitons instead of direct and indirect ones. That is, if the electron component of the exciton is a spectator while the hole component goes through a spin-conserving intervalley transition. In this case, however, the transition is governed by the large spin-split energy of the valence band ($\Delta_{v,0} \gg \Delta_{c,0}$).

\subsection{Parameters} \label{sec:parameters}
We have used $a=3.2~\AA$ for the triangular  lattice constant for both ML-WSe$_2$ and ML-MoSe$_2$ (distance between transition-metal atoms), leading to $K_0 = 1.3~\AA^{-1}$ for the wavenumber that connects the valley centers. In addition, we have assumed that $\epsilon_d(K_0) = 2.5$ in both materials, following DFT calculations of the non-local dielectric function in ML-MoS$_2$.\cite{Qiu_PRB16,Latini_PRB15}  The materials below and above the ML have weak influence on the non-local dielectric constant at these large wavenumber values. Other parameters needed to evaluate the exciton self-energy and absorption spectrum are listed below. 

\vspace{2mm}
\textit{Mass parameters}: The effective mass parameters at the edges of the conduction and valence bands are taken from DFT calculations following Ref.~[\onlinecite{Kormanyos_2DMater15}]. The effective masses in the top and bottom valleys of the conduction band in ML-WSe$_2$ (ML-MoSe$_2$) are $0.29m_0$ and $0.4m_0$ ($0.58m_0$ and $0.5m_0$), respectively. The effective masses in the top and bottom valleys of the valence band in ML-WSe$_2$ (ML-MoSe$_2$) are $0.36m_0$ and $0.54m_0$ ($0.6m_0$ and $0.7m_0$), respectively. The masses are used when we calculate the exciton states and their masses ($M_i$ and $M_d$). The charge neutrality of the exciton, smallness of exciton radius in ML-TMDs,\cite{Stier_NanoLett16} and similar ballpark effective masses of electrons and holes suggest that the exciton only weakly interacts with the polar ML. When we calculate plasmon quantities, on the other  hand, we further increase the band-edge effective masses of electrons or holes in the Fermi sea because of their Fr\"{o}hlich interaction with the lattice. The amount by which we increase the effective mass is found from matching the trion (charged exciton) binding energy to the empirical values.\cite{VanTuan_PRB18} Very good agreement is achieved when the polaron effect amounts to an increase of $\sim$17\% in the effective mass of charged particles in ML-WSe$_2$ and $\sim$25\% in ML-MoSe$_2$. The larger mass increase in ML-MoSe$_2$ stems from the larger Fr\"{o}hlich interaction in this material. \cite{Sohier_PRB16} 

\vspace{2mm}
\textit{Binding energies in charge neutrality conditions}: We have employed the stochastic variational method calculations and parameters from Ref.~[\onlinecite{VanTuan_PRB18}], and got the following binding energies of direct and indirect excitons in charge-neutrality conditions. $|E_d|$ and $|E_i|$ are 178 and 195~meV, respectively, in h-BN/WSe$_2$/h-BN. Their respective values are $|E_d|=203$~meV and $|E_i|=211$~meV in h-BN/MoSe$_2$/h-BN.  The result for the direct-exciton in encapsulated ML-WSe$_2$ is also available experimentally and matches the calculated value of $|E_d|$.\cite{Stier_PRL18} When calculating the absorption spectrum, we consider the band-gap energy at charge-neutrality conditions, such that the direct-exciton peak emerges at 1.725~eV in ML-WSe$_2$ and at 1.65~eV in ML-MoSe$_2$. These values are just reference energy levels. 

\vspace{2mm}
 \textit{Local-field effect parameters}: We have calculated the values of $\eta_c$ and $\widetilde{\eta}$ in Eqs.~(\ref{eq:eta}) and (\ref{eq:eta_prime}) by employing hydrogen-like $5d$ ($4d$) orbitals in tungsten (molybdenum). The calculation method is detailed in Ref.~[\onlinecite{VanTuan_arXiv19}], and it yields that $\eta_c \approx 0.2$ and $\eta_v \approx 0.47$ and $\widetilde{\eta} \approx 0.42$ in both materials. As a result, electrons in ML-TMDs generate intervalley collective excitations more effectively than holes because of the orbital composition of electronic states in the conduction and valence bands.\cite{VanTuan_arXiv19} We get that for the aforementioned parameters, $\alpha_{0}=1.35$ in ML-WSe$_2$ and $\alpha_{0}=1.8$ in ML-MoSe$_2$ for electron-doped conditions whereas  their respective values in hole-doped conditions are $\widetilde{\alpha}_0=0.58$ and $\widetilde{\alpha}_0=1.03$. The enhanced values in electron doping stems from the slower decay of $\mathcal{F}_c({q})$ compared with $\mathcal{F}_v({q})$ when $q$ is increased, governed by the different orbital compositions in the conduction and valence bands.\cite{VanTuan_arXiv19} In addition, $\mathcal{F}_{v}(\bm{q})$ oscillates between positive and negative values when $q$ is increased while $\mathcal{F}_{c}(\bm{q})$ is kept positive. As a result, the interference between various umklapp processes is constructive in Eqs.~(\ref{eq:eta}) vs destructive in Eq.~(\ref{eq:eta_prime}), indicating that higher-order umklapp processes $|\mathbf{K}_0 + \bm{G}| > K_0$ are more effective in enhancing the damping-free propagation range of intervalley plasmons in electron-doped conditions ($1/\eta_c > 1/\widetilde{\eta}$).

\vspace{2mm}
 \textit{Spin-splitting energy}:
The energy splitting between the top and bottom valleys in the  conduction band has a dominant contribution from spin-orbit coupling as well as contributions from long-wavelength and shortwave exchange interactions,\cite{VanTuan_arXiv19,Scharf_arXiv18} 
\begin{equation}\label{Eq:Delta_c}
\Delta_c =  |\Delta_{c,0}| + (1 - \delta_v)(1-c_0)\left( \frac{1}{2} - \eta_c \alpha_0 \right)\varepsilon_F \,.
\end{equation}
$\Delta_{c,0}$ is the spin-splitting energy in the conduction band due to spin-orbit coupling, and $\delta_v=1\,(0)$ for hole- (electron-) doped conditions. To evaluate $\Delta_v$, we exchange the indices $c \leftrightarrow v$ and replace $\alpha_0$ and $\eta_c$ with $\widetilde{\alpha}_0$ and $\widetilde{\eta}$. The valence band values for $\Delta_{v,0}$ are taken from DFT-based calculations, where $|\Delta_{v,0}|=427$~meV in ML-WSe$_2$ and 185~meV in ML-MoSe$_2$.\cite{Kormanyos_2DMater15} These values match very well the empirical energy difference between type-A and type-B excitons (optical transition from the top and bottom valleys of the valence band). The values of $|\Delta_{c,0}|$ in the conduction band were extracted by assuming that dark excitons have the same binding energies as the indirect ones (because the electron effective masses are the same in both cases: Dark excitons are formed when the electron and hole reside in the same valley but their spin configuration forbids optical transitions for out-of-plane propagating photons). Using the empirical value for the dark excitons: 40~meV below the neutral direct-bright exciton in ML-WSe$_2$,\cite{Robert_PRB17,Zhang_NatNano17,Zhou_NatNano17} and about the same energy as that of the neutral direct-bright exciton in ML-MoSe$_2$,\cite{Wang_PRL17} we have extracted the spin-orbit contribution to the spin-splitting energy from $|\Delta_{c,0}| = 40 - |E_i - E_d| = 23$~meV in ML-WSe$_2$ and $|\Delta_{c,0}| = |E_i - E_d| = 8$~meV in ML-MoSe$_2$.  

\subsection{Broadening} \label{sec:broadening}

The final piece in our model is the choice of broadening parameters. For the direct-exciton case in Eq.~(\ref{Eq:G}), we use~\cite{HaugKoch_Book}
\begin{equation}\label{Eq:br}
\Gamma(\hbar\omega)=\Gamma_1+\frac{\Gamma_2}{1+\exp\left[\left(E_{\mathrm{g}}(n)-\hbar\omega\right)/\Gamma_3\right]},
\end{equation}
where $\Gamma_1$ is broadening due to radiative decay and band-gap fluctuations of the ML because of charged defects in the substrate. $\Gamma_2$ and $\Gamma_3$ describe enhanced homogenous broadening when $\hbar\omega$ crosses into the continuum, $\hbar\omega>E_\mathrm{g}(n)$.\cite{HaugKoch_Book,Huang_PRB88,Honold_PRB89,Moody_NatComm15} The density-dependent band gap, $E_\mathrm{g}(n)$, is calculated through the screened exchange and Coulomb-hole correlation due to long-wavelength plasmons (see Appendix \ref{app:bgr} for details). We use $\Gamma_1=3$~meV, $\Gamma_2=30$~meV and $\Gamma_3=10$~meV in the simulations below.  In addition, due to the energy dependence of the broadening function, the absorption of photons with energies close to the band gap energy is strongly suppressed compared with the absorption of photons with energies far below the band gap. As a result, this energy-dependent broadening introduces an artificial redshift of up to 5~meV, when the density increases from 0 to 5$\times$10$^{12}$~cm$^{-2}$. To compensate for this small redshift, we add a small density-dependent blueshift to $|E_d|$ that keeps the peak position constant in the absorption spectrum when the charge density increases and when $\Sigma_\mathrm{s}=0$. This correction has no bearing on the many-body effects we study, and it is not needed if one employs an energy-independent broadening function instead of Eq.~(\ref{Eq:br}).

\begin{figure}[t]
\centering
\includegraphics*[width=8.6cm]{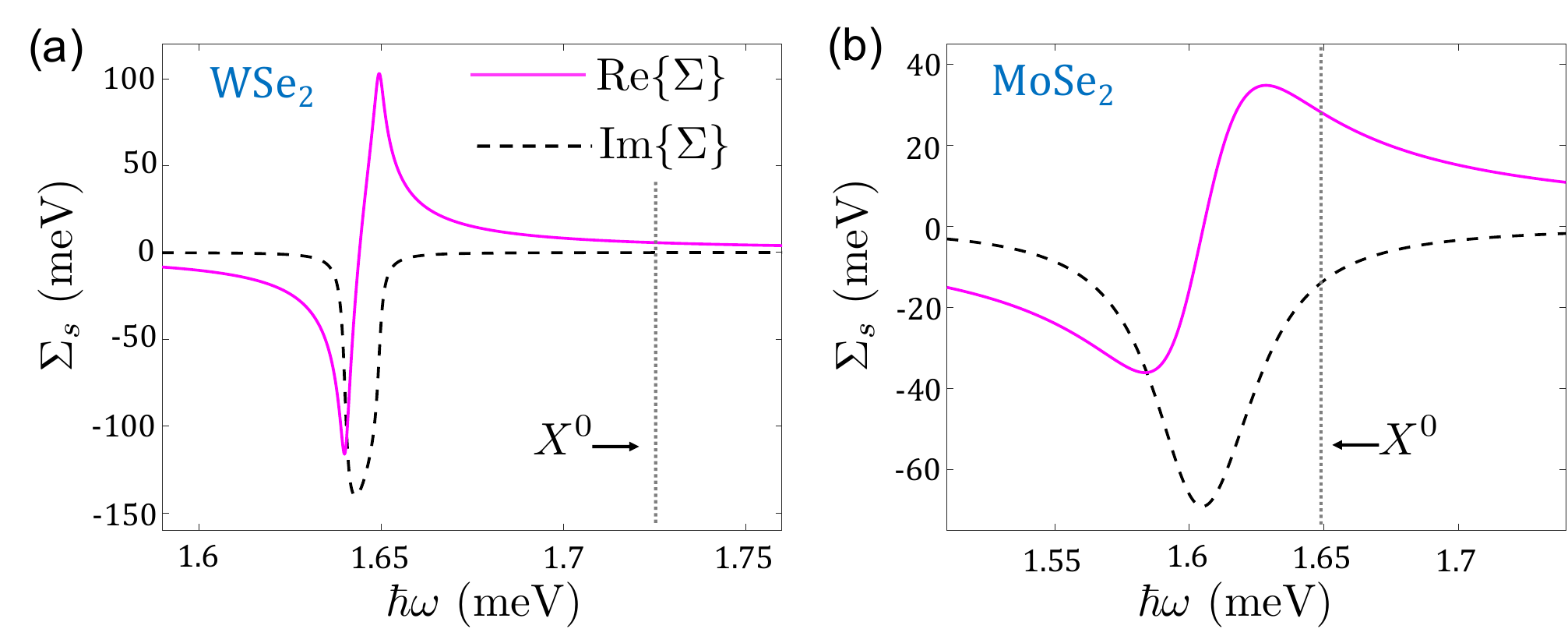}
\caption{The exciton self-energy in electron-doped ML-WSe$_2$ and ML-MoSe$_2$ at $n=5\times10^{12}$ cm$^{-2}$. The energy difference between the direct exciton ($X^0$) and the self-energy pole is larger in ML-WSe$_2$.}\label{fig:sigma}
\end{figure}

The broadening employed for the self energy function in Eq.~(\ref{Eq:SE}), $\Omega \rightarrow E + i\Gamma_{\Sigma}$, is dealt differently in ML-MoSe$_2$ and ML-WSe$_2$. A large broadening is needed in ML-MoSe$_2$ due to the energy proximity of direct and indirect excitons. In detail and using Fig.~\ref{fig:cartoon}(b) for guidance, the indirect exciton in ML-MoSe$_2$ is heavier than the direct one because $m_{ct}\approx 0.58m_0$ whereas  $m_{cb} \approx 0.5m_0$.\cite{Kormanyos_2DMater15} The resulting larger binding energy of the indirect exciton is offset by a larger band-gap energy, and consequently,  $E_{\mathrm{g},i} + E_i$ is close to $E_{\mathrm{g},d} + E_d$. Further support for this spectral overlap can be found from the absence of a spectrally resolved dark exciton in ML-MoSe$_2$.\cite{Wang_PRL17} In the context of our perturbative-based calculation, we use large broadening to avoid numerical instabilities in the renormalized Green's function when $E_{\mathrm{g},i} + E_i$ and $E_{\mathrm{g},d} + E_d$ are nearly degenerate. This problem does not arise in ML-WSe$_2$ [Fig.~\ref{fig:cartoon}(a)], where $m_{cb}\approx 0.4m_0$ and $m_{ct}\approx 0.29m_0$,\cite{Kormanyos_2DMater15} and as a result, $E_{\mathrm{g},i} + E_i$ is well below $E_{\mathrm{g},d} + E_d$. Indeed, experiments find that the dark-exciton energy is $\sim$40~meV below the bright one in ML-WSe$_2$.\cite{Robert_PRB17,Wang_PRL17,Zhou_NatNano17,Zhang_NatNano17,Barbone_NatComm18} Figure~\ref{fig:sigma} shows the calculated self energies in ML-WSe$_2$ and ML-MoSe$_2$ with $\Gamma_{\Sigma}=1$ and 20~meV, respectively.

\section{Results and Conclusions} \label{sec:res_conc}

\begin{figure*}[]
\centering
\includegraphics*[width=17cm]{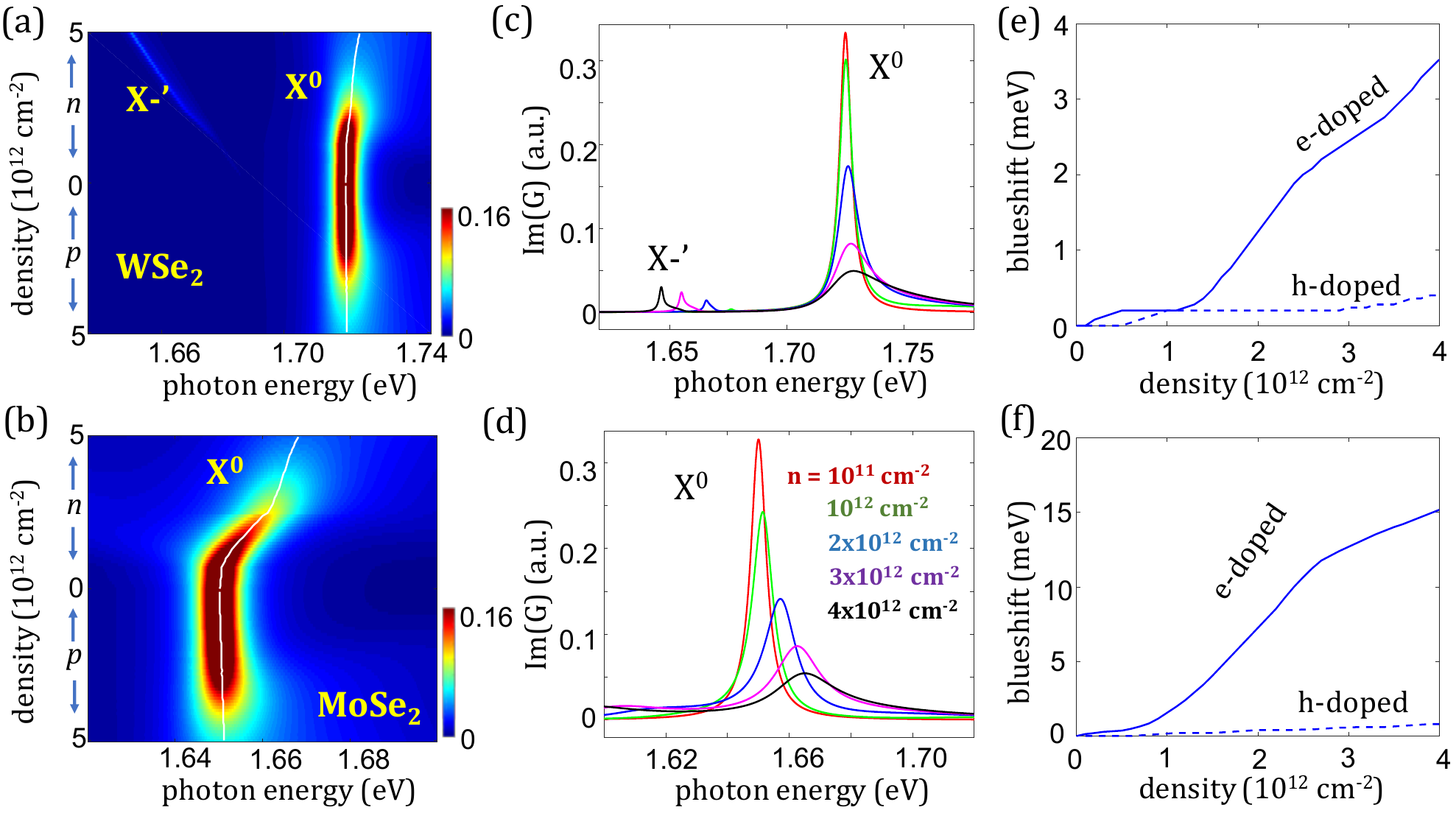}
\caption{Calculated absorption spectrum of the neutral exciton, $X^0$,  for ML-WSe$_2$ (top) and ML-MoSe$_2$ (bottom). (a,b) The absorption as functions of charge density and photon energy. The white lines trace the peak position. In addition, the exciton-plasmon interaction corresponds to $X-$' in the low-energy side of the spectrum in electron-doped ML-WSe$_2$. (c,d) Cross sections from panels (a) and~(b) for different electron densities. (e,f) The blueshift dependence of $X^0$ on charge density, where solid (dashed) lines denote electron (hole) doping. }\label{fig:Abs}
\end{figure*}

Figure~\ref{fig:Abs} shows the calculated absorption profile of neutral excitons, revealing good agreement with the experimental results in Fig.~\ref{fig:Exp}. The only free parameters pertain to broadening. The theory confirms that the blueshift of $X^0$ is observed only in electron-doped TMDs [Figs.~\ref{fig:Abs}(e) and (f)] and that it is larger in ML-MoSe$_2$ than in ML-WSe$_2$ [Figs.~\ref{fig:Abs}(c) and (d)]. The latter stems from the proximity between energies of direct and indirect excitons in ML-MoSe$_2$. The blueshift is weaker in hole-doped TMDs because of a smaller local-field effect and a mismatch between the plasmon energy when it is governed by $\Delta_{v,0}$ and the ten-fold smaller energy difference of direct and indirect excitons, governed  by $\Delta_{c,0}$  (Fig.~\ref{fig:cartoon}). Our analytical model captures the observed emergence and redshift of the optical sideband in ML-WSe$_2$ ($X-$'). The spectral position of this many-body feature is about one plasmon energy below the indirect exciton, which in WSe$_2$ lies at a lower energy than the direct exciton. A clear advantage of our theoretical model is its exceptional efficiency: All of the density-dependent many-body effects in Fig.~\ref{fig:Abs} are computed within seconds on a simple computer. 

In conclusion, we have measured the doping density dependence of the neutral-exciton energy shift in ML-TMDs. By using a transparent model, we can explain several many-body effects. While the competition between BGR and screening of the electron-hole interaction well describes the nearly constant position of the $X^0$ peak for hole doping, intervalley plasmons play a crucial role to describe $X^0$ in electron-doped samples. Renormalization of the pronounced $X^0$ absorption peak by these plasmons results in a blueshift with increasing doping density, which we can also observe experimentally. Ultimately, the strong exciton optical transitions in these materials will find use in a variety of optoelectronic applications,~\cite{Xu_NatPhys14,Britnell_Science13,Mak_NatPhot16,Lee_APL14,Lee_APL10,Sanchez_ACSNano14,Zutic_MT18} offering a wide range of wavelength tuning controlled by a gate voltage in van der Waals heterostructures. 

\acknowledgments{The work at the University of Rochester was supported by the Department of Energy, Basic Energy Sciences (Grant No. DE-SC0014349). The works at the University at Buffalo was by the Department of Energy, Basic Energy Sciences under Grant No. DESC0004890 (I.\v{Z}.), U.S. ONR N000141712793 (B.S.), and the German Science Foundation (DFG) Grant No. SCHA 1899/1-1 (B.S.). The work at the University of W\"urzburg was supported by the DFG Grant No. SFB 1170 ``ToCoTronics'' and by the ENB Graduate School on Topological Insulators. The work at the Pennsylvania State University was supported by the Department of Energy under Contract No. DESC0013883 (spectroscopy measurements), DESC0012635 (sample and device fabrication), and the National Science Foundation under Contract No. DMR-1410407 (Z.W).}


\appendix

\section{Band-gap renormalization}\label{app:bgr}
To calculate the broadening from Eq.~(\ref{Eq:br}), we need to evaluate the density-dependent shrinkage of the band-gap energy between the top valley in the valence band and bottom one in the conduction band. As the charge density is increased, the redshifting continuum renders the exciton more prone to scattering and dephasing processes. We model this effect through the enhanced broadening. The band-gap renormalization (BGR) is governed by the long-wavelength part of the dynamically-screened Coulomb potential, $W(\bm{q},\omega)$, where $qa \ll 1$ ($q$ is the wavenumber of the charge excitation and $a$ is the lattice constant).  Below we provide a brief summary of the BGR calculation. Interested readers can find a comprehensive analysis of this subject in Ref.~[\onlinecite{Scharf_arXiv18}].

The BGR has contributions from screened-exchange and Coulomb-hole energies. The former affects the populated bottom valley in the conduction band for electron-doped samples or populated top valley in the conduction band for holes-doped samples. The shift is largely the same for all of the low-energy states of the populated valleys, and therefore, we assume a rigid energy shift. The screened-exchange energy is simply half the Fermi energy in 2D systems\cite{Scharf_arXiv18}
\begin{equation}\label{Eq:sx_3}
\Sigma_{\mathrm{sx}} \approx  - \frac{1}{2}\varepsilon_{F} \,.
\end{equation}

\begin{figure}[t]
\includegraphics*[width=8.6cm]{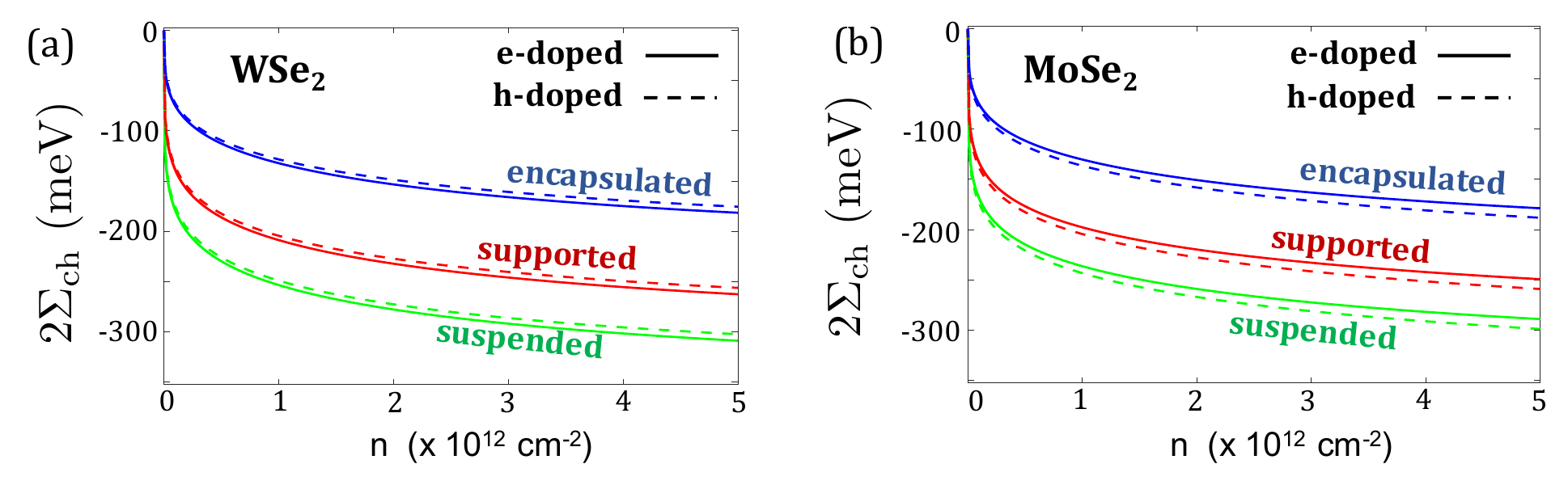}
\caption{The Coulomb-hole contribution to the band-gap renormalization. (a) and (b) show the results for ML-WSe$_2$ and ML-MoSe$_2$, respectively. Solid (dashed) lines correspond to electron- (hole-) doped samples. The results are shown for MLs encapsulated in h-BN, supported on SiO$_2$, and suspended in air. See Ref.~[\onlinecite{Scharf_arXiv18}] for further details.}\label{fig:ch}
\end{figure}

Next, we evaluate the Coulomb-hole energy due to long-wavelength plasma excitations, which is by far the dominant contribution to the BGR. The term Coulomb-hole refers to the lack of charge next to a charged particle due to the Pauli exclusion principle. Unlike the screened-exchange contribution, the Coulomb-hole energy is largely the same for populated and unpopulated valleys. Conduction bands shift down in energy while valence bands shift up, and the shift has similar magnitude in ML-TMDs. The Coulomb-hole energy reads\cite{Scharf_arXiv18}
\begin{eqnarray}\label{eq:ch4}
\!\!\!\!\!\! \!\!\!\!\!\! \Sigma_{\mathrm{ch}} \approx  - \frac{ e^2 }{2} \int_0^{q_c} \! \!   \frac{dq}{\epsilon_d(q)} \! \cdot \! \left[ 1 \!+\! \frac{q}{\kappa(q)} \!+\! C_\mathrm{eff} \!\left(\! \frac{\varepsilon_{b,\bm{q}}}{\omega_{\ell}(q)}\!\right)^{^2}  \right]^{-1}\!\!\!\!\!\!\!,
\end{eqnarray}
The integration cutoff, $q_c$, denotes the fact that plasmons whose energy is much larger than the Fermi energy experience Landau damping due to single-particle excitations. $C_\mathrm{eff}$ is a constant of the order of unity needed to compensate for the fact that the static approximation from which we have calculated the Coulomb-hole energy typically overestimates the screening effect. $\hbar\omega_{\ell}(q)$ is the energy of 2D plasmons in the long-wavelength limit,\cite{HaugKoch_Book}
\begin{eqnarray}\label{Eq:lwplasmon}
\hbar\omega_{\ell}(q)=\sqrt{\frac{2e^2\varepsilon_F q}{\epsilon_d(q)}}\,\,,
\end{eqnarray}
and the parameter $\kappa(q)$ in Eq.~(\ref{eq:ch4}) is the screening length calculated from the static limit of the RPA dielectric function,\cite{Ando1982:RMP} 
\begin{equation}\label{eq:kappa_RPA}
\!\!\! \kappa(q)\! = \! \frac{g_sg_v e^2 m^{\ast}}{\hbar^2\epsilon_d(q)} \! \left[1-\sqrt{1\!-\!\left( \! \frac{2k_\mathrm{F}}{q}\!\right)^2}\Theta(q\!-\!2k_\mathrm{F})\right]\!\!.\,\,\,\,\,\,
\end{equation}
$g_s=1$ and $g_v=2$ are the spin and valley degeneracies, respectively. $m^{\ast}=m_{cb}$ in electron-doped samples, while $m^{\ast}=m_{vt}$ hole-doped samples.  Finally,  $\epsilon_d(q)$ in Eqs.~(\ref{eq:ch4}), (\ref{Eq:lwplasmon}), and (\ref{eq:kappa_RPA}) is the non-local dielectric function in the long-wavelength limit. This function can be described through the Rytova-Keldysh potential of a thin semiconductor,\cite{Rytova_MSU67,Keldysh1979:JETP,Cudazzo2011:PRB,Trolle_SR17,Meckbach_PRB18} first-principles calculations,\cite{Qiu_PRB16,Latini_PRB15} or a model that considers a ML-TMDs as a system made of three atomic sheets.\cite{VanTuan_PRB18} All of these non-local dielectric function models converge to the value given by the average effective dielectric constants of the materials below and above the ML when $q \rightarrow 0$. Therefore, they yield qualitatively similar results for the Coulomb-hole energy. Here, the non-local dielectric function and its parameters are taken from Ref.~[\onlinecite{VanTuan_PRB18}].

All in all, we get that $E_g(n)$ in Eq.~(\ref{Eq:br}) follows
\begin{equation}\label{eq:eg0}
E_g(n) = E_{g,0} + 2\Sigma_{\mathrm{ch}} + \Sigma_{\mathrm{sx}}\,,
\end{equation}
where $E_{g,0}$ is the  reference  level for the band-gap energy  at vanishing densities. Figure~\ref{fig:ch} shows the Coulomb-hole contribution to the band-gap renormalization, $2\Sigma_{\mathrm{ch}}$,  where the factor of 2  comes from the simultaneous energy downshift and upshift of the conduction and valence bands, respectively.  We have used $C_\mathrm{eff}=4$ and $\hbar^2q_c^2/2m_b=0.12$~eV in all of the calculations.

\end{document}